\documentclass[12pt,preprint]{aastex}
\usepackage{graphicx, amsmath, amssymb}
\usepackage[dvips]{color}
\usepackage{lineno} 

\newcommand{\pow}[2]{\ensuremath{\mbox{#1}^{\rm #2}}}
\newcommand{\subs}[2]{\ensuremath{\mbox{#1}_{\rm #2}}}
\newcommand{\del}[1]{\ensuremath{\Delta #1}}
\newcommand{\Blos}{\ensuremath{B_{\rm los}}}

\newcommand{\methanol}{CH$_3$OH}
\newcommand{\kmS}{km \pow{s}{-1}}
\newcommand{\mG}{\pow{mG}{-1}}

\shortauthors{Momjian \& Sarma}
\shorttitle{Two epochs of the Zeeman Effect in the 44 GHz CH$_3$OH maser in OMC-2}

\begin{document}

\title{Comparison of Two Epochs of the Zeeman Effect in the \\
	44 GHz Class I methanol (CH$_3$OH) maser line in OMC-2}

\author{E.\ Momjian\altaffilmark{1}, A.~P.\ Sarma\altaffilmark{2}}

\altaffiltext{1}{National Radio Astronomy Observatory, Socorro NM 87801; emomjian@nrao.edu}

\altaffiltext{2}{Physics Department, DePaul University, 
2219 N. Kenmore Ave., Byrne Hall 211, 
Chicago IL 60614; asarma@depaul.edu}

\begin{abstract}
We present a second epoch of observations of the 44 GHz Class I methanol maser line toward the star forming region OMC-2. The observations were carried out with the Very Large Array, and constitute one of the first successful Zeeman effect detections with the new WIDAR correlator. Comparing to the result of our earlier epoch of data for this region, we find that the intensity of the maser increased by 50\%, but the magnetic field value has stayed the same, within the errors. This suggests that the methanol maser may be tracing the large-scale magnetic field that is not affected by the bulk gas motions or turbulence on smaller scales that is causing the change in maser intensity.

\end{abstract}


\section{INTRODUCTION}
\label{sINTRO}


The ability to carry out high angular resolution observations of several types of masers has increased tremendously due to the availability of interferometers at several different frequencies. For example, 36 GHz methanol maser observations are now routine with the Karl G.~Jansky Very Large Array (VLA), e.g., \citet{sjouwerman10} and \citet{fish11}. Since masers allow for the observation of magnetic fields at high angular resolution via the Zeeman effect, there has often been speculation on whether they trace the large-scale magnetic field or only the field within the small-scale environments in which such masers form (\citealt{fr2006}; \citealt{wv2010}; \citealt{sm2009,sm2011}). Since the observation of magnetic fields is challenging in general (e.g., \citealt{tt2008}), the ability to measure the large-scale magnetic field with masers would prove to be invaluable, especially because magnetic fields play such an important role in the process of star formation (e.g., \citealt{mckee2007}; \citealt{bp2007}). 

At a distance of 450~pc (\citealt{gs1989}), the Orion Molecular Cloud 2 (OMC-2) is considered one of the nearest intermediate-mass star forming regions (e.g., \citealt{taka2008}). 
This region is located at an angular distance of 12$\arcmin$ northeast of the Trapezium OB cluster, and along with
OMC-3, it is part of a single long filament in Orion (\citealt{cas1995,chini97}).
In 2011, we reported the discovery of the Zeeman effect in the 44 GHz Class I \methanol\ maser line toward OMC-2 (\citealt{sm2011}). In this paper, we report follow-up observations on the Zeeman effect of this methanol maser line. These observations provide the opportunity to study whether the intensity or the magnetic field has changed since our first observations (\citealt{sm2011}), and the implications this would have for measuring magnetic fields with 44 GHz \methanol\ masers.
The observations and data reduction are presented in \S \ref{sODR}. In \S \ref{sANAL}, we present the analysis to derive the line of sight magnetic field value through the Zeeman effect. Finally, we present the results and the discussion
in \S\ \ref{sR}.

\section{OBSERVATIONS, DATA REDUCTION}
\label{sODR}

We observed the $7_{0}-6_1\, A^+$ \methanol\ (methanol) maser emission line at 44 GHz with the VLA\footnote{The National Radio Astronomy Observatory (NRAO) is a facility
of the National Science Foundation operated under cooperative agreement by Associated Universities, Inc.}
on 2011 September 25. The observations were carried out in dual polarization with a 1 MHz
bandwidth and 256 spectral channels.
At the time of the observations, the array was in the later stages of the
reconfiguration from the most extended, A-configuration (maximum baseline = 36~km), to the most compact,
D-configuration (maximum baseline = 1~km). On the day of the observations, all antennas except for one were
already placed in D-configuration. The sole unmoved antenna that resulted in baselines $\gg 1$~km was
edited out during the data reduction process. The total observing time was 4~hours.
The calibrator source J0542$+$4951 (3C147) was used to derive the
antenna-based amplitude gain factors to calibrate the flux density scale.
The uncertainty in the flux density calibration at the observed frequency, accounting for various
observational parameters (e.g., weather, reference pointing, elevation effects), is expected to be up to 10\%.
Table\ \ref{tOP}\ summarizes the parameters of these observations.

Data reduction, including calibration, deconvolution and imaging, was carried out 
using the Astronomical Image Processing System (\citealt{greisen03}).
The flux density scale of the target source was set by applying the 
amplitude gain calibration solutions of 3C147. After Doppler correcting the data of OMC-2,
the spectral channel with the brightest maser emission signal was split off,
and self-calibrated in both phase and amplitude and imaged
in a succession of iterative cycles.
The final self-calibration solutions were 
applied to the full spectral-line data set of the target source.
Hanning-smoothed Stokes $I$=(RCP$+$LCP)/2 and $V$=(RCP$-$LCP)/2\footnote{RCP is right- and 
LCP is left-circular polarization incident on the antennas.
RCP has the standard radio definition of clockwise rotation of the 
electric vector when viewed along the direction of wave propagation.} 
image cubes were then constructed with a synthesized beamwidth of $2.17\arcsec \times 1.96\arcsec$,
and an effective velocity resolution of 0.053 \kmS.
The resulting image cubes were further processed using the MIRIAD data reduction package
to derive the magnetic field value.

We note that these observations are one of the first successful Zeeman effect observations and detection with the  Wide-band Digital Architecture (WIDAR) correlator. In 2010, the VLA/WIDAR commissioning team discovered a small but significant unanticipated spectral response, or ``spectral splatter'' effect while observing strong spectral lines at high frequency resolutions. Zeeman effect studies, which in this case rely on the difference of two strong signals, were particularly vulnerable. The ``spectral splatter'' problem resulted in corrupt Stokes $V$ spectra, and sometimes in false Zeeman effect detections, e.g., in the strongest 36~GHz methanol maser line in the star forming region DR21W \citep{fish11,MSF12}. This rendered the S-curve profiles in the Stokes\,$V$ data very unreliable \citep{sault10,sault12a,sault12b}. While the various Zeeman effect commissioning tests were concluded in April 2012, these tests showed that spectral line data acquired since August 2011 with WIDAR would be suitable for Zeeman effect and other spectral-line polarization studies. The OMC-2 observations reported in this paper were carried out in light of the various changes that have been implemented in the VLA/WIDAR system to address the spectral splatter problem. The data have been checked carefully to verify that the detected signal in Stokes V is astronomical in nature and not instrumental.

\section{ANALYSIS}
\label{sANAL}

Following the arguments presented in \S 3 of \citet{sm2011} for maser lines with widths  \del{\nu}  much greater than
the Zeeman splitting \subs{\del{\nu}}{z}, we measured the Zeeman effect 
by fitting the Stokes $V$ spectra in the least-squares sense to the equation   
\begin{equation}
\mbox{V}\ = \mbox{aI}\ + \frac{\mbox{b}}{2}\ 
\frac{\mbox{dI}}{\mbox{d}\nu}
\label{eVLCO}
\end{equation}
(\citealt{th82}; \citealt{skzl90}), where $a$ and $b$ are the fit parameters.
Here, $a$ is usually the result of small
calibration errors in RCP versus LCP, and it is on the
order of $10^{-4}$ or less in our observations.
The magnetic field value is obtained through the fit parameter $b$, where $b = zB$\,cos\,$\theta$.
Here, $z$ is the Zeeman splitting factor (Hz \mG), $B$ is the magnetic field, and $\theta$ is the
angle of the magnetic field to the line of sight (\citealt{ctg93}). The value of the Zeeman
splitting factor $z$ for \methanol\ masers is not known but is likely to be very small because \methanol, like H$_2$O, is a non-paramagnetic molecule. Therefore, we will report only values of $z\Blos$. 
We note that while eq.~(\ref{eVLCO})\ is 
true only for thermal lines, numerical solutions of radiative transfer equations
(e.g., \citealt{nw92}) have shown that it also results in reasonable values for the magnetic fields in masers.

\section{RESULTS AND DISCUSSION}
\label{sR}

\begin{figure}
\centering
\plottwo{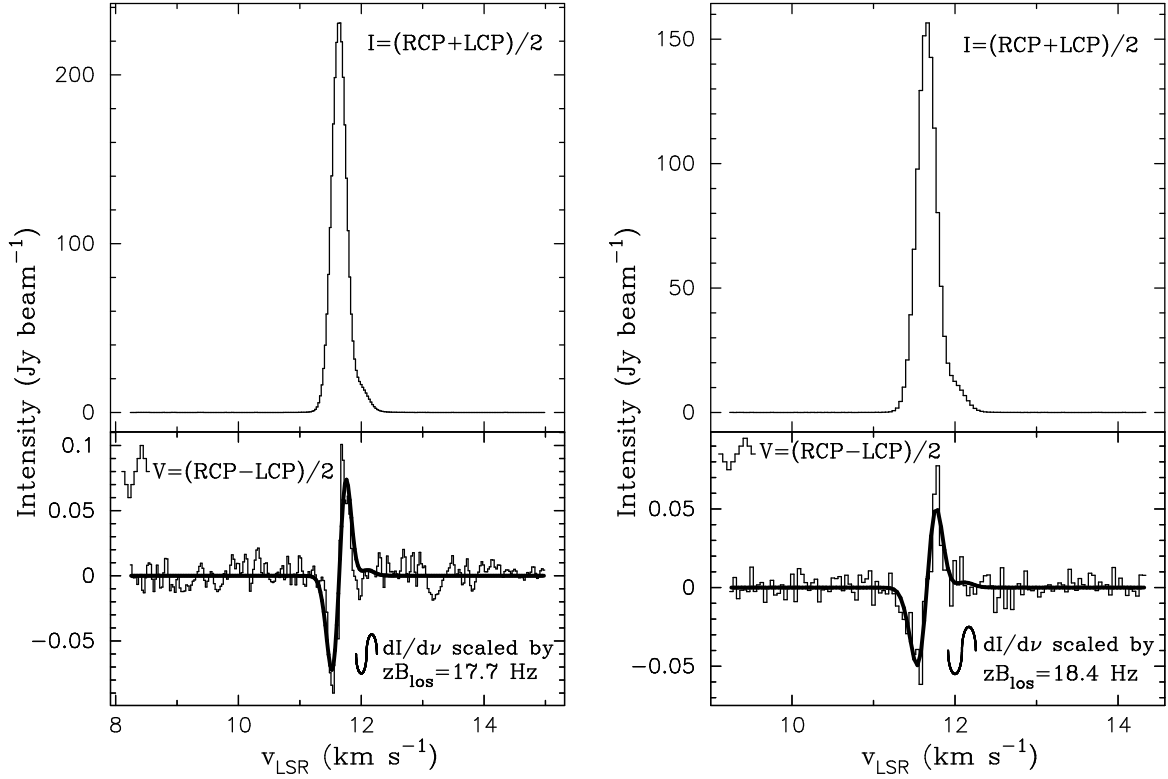}{./f1.eps}
\caption{Stokes $I$ ({\em top$-$histogram}) and $V$ ({\em bottom$-$histogram})
profiles of the maser toward the 44 GHz Class I \methanol\ maser in OMC-2. The left panel shows the data from our current observations, and the right displays the data from \citet{sm2011}. 
The curve superposed on $V$ in the lower frame is the derivative of $I$ scaled by a value of $z\Blos\ = 17.7 \pm 0.9$~Hz in the left panel and $z\Blos\ = 18.4 \pm 1.1$~Hz in the right panel respectively.
\label{fIVD}}
\end{figure}

Figure \ref{fIVD}\ (left panel) shows the Stokes $I$ and $V$ spectra of the 44 GHz \methanol\ maser toward OMC-2 from our current observations. As described in \S\ \ref{sANAL}, we determined magnetic fields by fitting the Stokes $V$ spectra in the least-squares sense using equation (\ref{eVLCO}). The fit parameter $b=z$\Blos\ (see eq.~\ref{eVLCO}) obtained by this procedure is $17.7 \pm 0.9$~Hz. By convention, a positive value for \Blos\ indicates a field pointing away from the observer. 

Also shown in Figure \ref{fIVD}\ (right panel) is the Stokes $I$ and $V$ spectra of the same 44 GHz \methanol\ maser from \citet{sm2011}. We conclude that this is the same maser as in the follow-up observations because the masers in both epochs have the same velocity ($v_\text{LSR}$), and are at the same position relative to other, weaker masers in the field. \citet{sm2011} found for this maser that $b = z\Blos = 18.4 \pm 1.1$ Hz.
Compared to \citet{sm2011}, the intensity of the maser is a factor of 1.5 higher in the current observations, while the values of $z$\Blos\ are the same, within the errors.

In order to obtain the value of \Blos\ from $b$, we need to know the Zeeman splitting factor $z$.
An empirical value for the Land\'{e}
$g$-factor can be obtained by extrapolation from lab measurements of several methanol maser transitions near 25 GHz made by \citet{jen51}.
Using this $g$-factor would result in a value of $z$
for the 44 GHz methanol maser line of 0.1~Hz~mG$^{-1}$, implying a line-of-sight
magnetic field value of order 0.2~G. Such a value is an order of magnitude higher
than those measured by water masers in star forming regions
(e.g., \citealt{sm2002,sm2008}). Moreover, the empirical scaling relation 
between the hydrogen molecule number density and magnetic field, $B \propto n^{0.47}$ \citep{rc1999},
would result in a number density of $10^{11}$~cm$^{-3}$, a value that is several
orders of magnitude higher than that thought to be suitable for the 44 GHz methanol
maser line (e.g., \citealt{prat2008}). \citet{wv2011} presented a detailed
discussion on the uncertainty of the $g$-factor derived by \citet{jen51} while reporting
the magnetic field values of various 6.7~GHz methanol masers. These authors concluded that the $g$-factor may be uncertain by an order of magnitude. 
Due to this, we have chosen to leave our results in terms of $b$ rather than \Blos.
Our observations and
the above noted considerations motivate the immediate need for an experimental measurement of
the Land\'{e} $g$-factor in the 44 GHz methanol maser line.

With the detection of consistent Stokes V signals in the 44~GHz methanol maser line in OMC-2 through two observing epochs, it is timely to assess whether the detected signal can be attributed to a magnetic field, and therefore being due to a real Zeeman effect.
As described by \citet{wv2011}, a potential effect that could mimic a regular Zeeman effect is caused by a rotation of the axis of symmetry for the molecular quantum states, and can occur when the stimulated emission rate of the maser $R$ becomes greater than the Zeeman splitting $g\Omega$ as the brightness of the maser increases while it becomes more saturated.
The detected splitting in our two epoch observations (this work and \citealt{sm2011}) is $g\Omega \simeq 18$~s$^{-1}$. The value of $R$ can be estimated using $R\simeq A k T_{\rm b} \Delta \Omega / 4 \pi h \nu$ (e.g., \citealt{wv2011}), where $A$ is the Einstein coefficient for the 44~GHz methanol maser transition, which is $0.392 \times 10^{-6}$~s$^{-1}$ \citep{cr93}, $k$ and $h$ are the Boltzmann and Planck constants, respectively, $\nu$ is the frequency of the maser transition, $T_{\rm b}$ is the maser brightness temperature, and $\Delta \Omega$ is the maser beaming solid angle. Higher angular resolution observations of the brightest 44~GHz methanol maser in OMC-2 resulted in $T_{\rm b} = 1.74 \times 10^7$~K and $\Delta \Omega = 4.8\times 10^{-3}$ \citep{sk2009}. From these parameters, we derive $R \sim 10^{-3}$~s$^{-1}$. This is several orders of magnitude lower than $g\Omega$, and suggests that the detected Stokes V signal is unlikely to be due to this non-Zeeman effect. Furthermore, this particular effect would introduce maser-intensity-dependent circular polarization because the molecules interact more strongly with the radiation field than with the magnetic field \citep{wv2011}. Since we observe no change in splitting over two epochs (\citealt{sm2011} and this work) in spite of the maser intensity having increased by a factor of 1.5, this is additional proof that the detected signal is due to the Zeeman effect in a magnetic field.


As noted above, the intensity of the maser is a factor of 1.5 higher in the current observations compared to \citet{sm2011}. Very few observations exist of the variability of 44 GHz Class I methanol masers, or indeed of any kind of Class I methanol maser. Such masers are believed to be excited in outflows in star forming regions. \citet{kurtz2004}\ observed some variability in a number of sources, but due to differences in spatial and spectral resolution and coverage, were able to confirm variability in only two 44 GHz methanol maser sources. More recently, \citet{pp2007}\ have reported short term variability on the scales of days or possibly even hours, in 44 GHz Class I masers toward the DR21 region.

Even though the 44 GHz methanol maser intensity in OMC-2 has increased, the measured magnetic field (as given by $z\Blos$) has stayed the same, within the errors. This leads us to speculate that the mechanism responsible for the change in maser intensity may be active on smaller scales than the shock compression that sets up the magnetic field. To elaborate, the post-shock magnetic field that we are detecting via the Zeeman effect was likely amplified above its pre-shock value in proportion to the density of the post- and pre-shock regions (e.g., equation (2) of \citealt{sm2011}). This does not appear to have changed between the two epochs, as $b = z\Blos$ has remained the same. Meanwhile, variability in collisionally pumped masers such as the 44 GHz Class I methanol maser likely results from bulk gas flows and turbulence moving parts of the maser column in and out of velocity coherence along the line of sight (\citealt{gray2005}). Evidently, a larger column of gas has been moved into the line of sight to increase the maser intensity, but the compression factor between post- and pre-shock regions has remained the same, thereby keeping the magnetic field the same. It would be very interesting, therefore, to observe other regions and compare maser intensities and magnetic fields between epochs in order to determine how frequently this happens. If the effect were widespread, it might mean that we are detecting the large-scale magnetic field in the shocked region that is not affected by any changes on smaller scales that cause the intensity of the maser to change.

\section{CONCLUSIONS}
\label{sCONC}

We have presented a second epoch of observations of the 44 GHz Class I methanol maser line toward the star forming region OMC-2 with the VLA. We detected a magnetic field value of $z$\Blos\  $= 17.7 \pm 0.9$ Hz. The observations constitute one of the first successful Zeeman effect detections with the new WIDAR correlator of the VLA. Comparing to the result of our earlier epoch of data for this region \citep{sm2011}, we find that the intensity of the maser has changed by a factor of 1.5, but the magnetic field value has stayed the same, within the errors. We speculate that we may be detecting the large-scale magnetic field in the post-shock region that is unaffected by bulk motions or turbulence on smaller scales that may be causing the change in the maser intensity.

\begin{deluxetable}{lccccccrrrrrr}
\tablenum{1}
\tablewidth{0pt}
\tablecaption{Parameters for VLA Observations \protect\label{tOP}}   
\tablehead{
\colhead{Parameter}  &  
\colhead{Value}}
\startdata
Observation Date & 2011 Sep 25 \\
Configuration &   D \\
R.A.~of field center (J2000) & $05^\text{h} 35^\text{m} 27^\text{s}.66$ \\
Decl.~of field center (J2000) & $-$05$\arcdeg$09$\arcmin$39$\arcsec$.6  \\
Total Bandwidth & 1 MHz \\
No. of channels & 256 \\
Total Observing Time  & 4 hr \\
Rest Frequency & 44.069488 GHz \\
Target source velocity & 11.6 \kmS \\
Hanning Smoothing &  Yes \\
Effective velocity resolution & 0.053  \kmS \\
FWHM of synthesized beam & $2.17\arcsec \times 1.96\arcsec$  \\
& P.A. $= 15\arcdeg$ \\
Line rms noise\tablenotemark{a} &   10 mJy beam$^{-1}$
\enddata
\tablenotetext{a}{The line rms noise was measured from the stokes $I$ 
image cube using maser line free channels.}
\end{deluxetable}
     
\clearpage

\end{document}